\documentclass[twocolumn,showpacs,preprintnumbers,amsmath,amssymb]{revtex4}
\usepackage{graphicx}% Include figure files
\usepackage{dcolumn}% Align table columns on decimal point
\usepackage{bm}% bold math

\begin{document}

\title{Duality, Quantum Skyrmions and the Stability of an SO(3)
                      2D Quantum Spin-Glass \\}
%\\with Forced Linebreak% Force line breaks with \\

\author{  C.M.S. da Concei\c c\~ao$^1$ and E.C.Marino$^{2}$}
\affiliation{$^1$Departamento de F\'\i sica Te\'orica, Universidade
do Estado do Rio de Janeiro, Rio de Janeiro, RJ 20550-013, Brazil
 }
 %\altaffiliation[2 ]{Department of Physics, Princeton University, Princeton, NJ 08544, USA}%Lines break automatically or can be forced with \\
%\author{}%
\affiliation{$^2$Instituto de F\'\i sica, Universidade Federal do
Rio de Janeiro, Cx.P. 68528, Rio de Janeiro, RJ 21941-972, Brazil}

%\author{E.C. Marino}
%\affiliation{
%Department of Physics, Princeton University, Princeton, NJ 08544, USA\\

\date{\today}% It is always \today, today,

\begin{abstract}

Quantum topological excitations (skyrmions) are analyzed from the
point of view of their duality to spin excitations in the different
phases of a disordered two-dimensional, short-range interacting,
SO(3) quantum magnetic system of Heisenberg type. The phase diagram
displays all the phases, which are allowed by the duality relation.
We study the large distance behavior of the two-point correlation
function of quantum skyrmions in each of these phases and, out of
this, extract information about the energy spectrum and
non-triviality of these excitations. The skyrmion correlators
present a power-law decay in the spin-glass(SG)-phase, indicating
that these quantum topological excitations are gapless but
nontrivial in this phase. The SG phase is dual to the AF phase, in
the sense that topological and spin excitations are respectively
gapless in each of them. The Berezinskii-Kosterlitz-Thouless
mechanism guarantees the survival of the SG phase at $T \neq 0$,
whereas the AF phase is washed out to $T=0$ by the quantum
fluctuations. Our results suggest a new, more symmetric way of
characterizing a SG-phase: one for which both the order and disorder
parameters vanish, namely $\langle \sigma \rangle = 0 $, $\langle
\mu \rangle =0 $, where $\sigma$ is the spin and $\mu$ is the
topological excitation operators.

\end{abstract}

\maketitle

\section{\label{sec:level1}Introduction}

\renewcommand{\theequation}{\arabic{section}.\arabic{equation}}
\setcounter{equation}{0}

Given a physical system, the spectrum of its possible excitations
can, in principle, be roughly divided into two groups: hamiltonian
and topological excitations. The first ones correspond to states,
which are created out of the ground state by the action of operators
that appear explicitly in the hamiltonian. These excitations carry
in general quantum numbers corresponding to physical quantities,
such as charge for instance, which are conserved as a consequence of
some continuous symmetry of the hamiltonian. Topological
excitations, conversely bear quantum numbers that correspond to
quantities whose conservation derives from a nontrivial topology of
the space of classical configurations and whose continuity equation
is satisfied as an identity rather then being the result of some
continuous symmetry in the system.

In this work, we investigate the correlation functions of quantum
skyrmions, which are the topological excitations that may occur in
two-dimensional magnetic systems described by a hamiltonian of the
Heisenberg type, with SO(3) symmetry and
nearest-neighbors-interactions on a square lattice. We consider the
quenched disordered case, with a Gaussian random distribution of
exchange couplings centered at an antiferromagnetic (AF) coupling
$\bar J > 0$, with variance $\Delta J$.

This disordered system has been studied by means of a mapping onto a
generalized nonlinear sigma model (NLSM),
 in which the original (staggered) spin
is mapped onto the NLSM field $\mathbf{n}^{\alpha}$
\cite{mm,mmartigo} and the index $\alpha = 1,...,n$ corresponds
  to the different replicas which are required for a quenched
  average. Eventually we must take the limit $n\rightarrow 0$ \cite{ea,by,mpv}.

A remarkable feature of this system is that, whereas the quantum
fluctuations completely ``wash out'', the ordered N\'eel phase at $T
\neq 0$, in agreement with the Mermin-Wagner theorem \cite{mw}, the
spin glass phase persists even at a finite temperature. Even though
this does not violate the theorem, because there is no spatial long
range order in the SG phase, it is somewhat intriguing to have a SG
phase at a finite temperature in a quantum 2D system, specially if
we consider a lot of evidence against the occurrence of a SG phase
at a finite temperature in 2D Ising systems \cite{ising}. We will
see below, however, as a consequence of our study of quantum
topological correlation functions, that this fact can be understood
as a manifestation of the Berezinskii-Kosterlitz-Thouless (BKT)
mechanism \cite{bkt,mond}. Indeed, it is the BKT mechanism that
allows the existence of a SG phase at a finite temperature in this
2D system. Moreover, the action of this mechanism is only possible
in the case of SO(3) systems, where a vortex picture for the
skyrmions does exist. This explains why the corresponding SG phase
is not found in 2D Ising sytems, where such a picture is absent.

\section{\label{sec:level1}Duality and Quantum Topological Excitations}

\renewcommand{\theequation}{\arabic{section}.\arabic{equation}}
\setcounter{equation}{0}

 The non-triviality of the
topology of classical configurations space, is ultimately
responsible for the stability of classical topological excitations.
At the quantum level, this nontrivial topology manifests as a
degeneracy of the ground state. The existence of ground-state
degeneracy, therefore, is the indication that, at a
quantum-mechanical level, the system presents quantum topological
excitations in its spectrum. These excitations, differently from the
former, cannot be created by acting on the ground state with
hamiltonian operators. A familiar example of topological excitations
are magnetic vortices in 2D, which occur in type II superconductors.
The corresponding topological ``charge'' would be the magnetic flux
piercing the plane.

Topological excitations, surprisingly, are related to the concept of
order-disorder duality, which plays an important role in many areas
of physics. This can be seen in the following way. Consider a system
characterized by a dynamical variable $\sigma$. It could be, for
instance, the magnetization on a ferromagnetic system or the
staggered magnetization in an antiferromagnetic one. Suppose the
hamiltonian $H[\sigma]$ possesses a symmetry, which at the quantum
level is implemented by an unitary operator $U$ such that, for $g$
being the element of the symmetry group, $U \sigma U^\dagger = g
\sigma$, with $[H,U]=0$. Since
\begin{equation}
\langle \sigma \rangle = \langle 0| \sigma |0\rangle = g \langle 0|
U^\dagger \sigma U|0\rangle, \label{1}
\end{equation}
it follows that if $\langle \sigma \rangle \neq 0$, then necessarily
$U|0\rangle = |0'\rangle \neq |0\rangle$, that is, the ground-state
will be degenerate. This, however, as remarked above, is a sign of
non-triviality of the topological excitations. Now, assume the
one-particle quantum topological excitation state is given by
$|\mathrm{Top}\rangle =\mu |0\rangle $, where $\mu$ is the operator
that creates these excitations out of the ground state. Then, if
these states are nontrivial, we must have $\langle 0|\mu |0\rangle
=0 $ because this means $|\mathrm{Top}\rangle $ is orthogonal to the
ground state, as a genuine excited state must be. We conclude,
therefore, that $\langle \sigma \rangle \neq 0$ implies $\langle \mu
\rangle = 0$.

Conversely if  $\langle 0|\mu |0\rangle \neq 0 $, this would mean
that $|\mathrm{Top}\rangle $ is not orthogonal to the ground state
and hence is actually not a genuine excitation. This would imply the
ground state should be unique, because otherwise topological
excitations should exist as nontrivial states. Now, if the ground
state is unique, we have $U|0\rangle = |0\rangle$ and (\ref{1})
would imply $\langle \sigma \rangle =0$. We therefore conclude that
$\langle \mu \rangle \neq 0$ implies $\langle \sigma \rangle = 0$.

The previous analysis shows that, if $\langle \sigma \rangle $
measures the amount of order (an order parameter) then $\langle \mu
\rangle $ measures the amount of disorder, being naturally a
disorder parameter. We see that a duality relation exists between
the topological excitations creation operator $\mu$ and the
hamiltonian operator $\sigma$. The physical and mathematical
properties of this duality relation are captured by the so-called
dual algebra,
 satisfied by $\sigma$ and $\mu$ \cite{da}, namely
\begin{equation}
 \mu(\mathbf{x}, t)  \sigma (\mathbf{y}, t)  = g(\mathbf{y} -\mathbf{x})  \sigma (\mathbf{y}, t) \mu(\mathbf{x}, t),
\label{2}
\end{equation}
where, for each fixed $\mathbf{x}$ and $\mathbf{y}$, $g(\mathbf{x}
-\mathbf{y})$ is an element of the symmetry group. This relation is
the basis for constructing the topological excitations creation
operator $\mu$. It implies, for instance,
\begin{equation}
\langle \sigma \rangle  \langle \mu \rangle  = 0, \label{22}
\end{equation}
hence we cannot have both $\langle \sigma \rangle$ and $\langle \mu
\rangle$ nonvanishing. It also implies the spectrum is gapless
whenever both $\langle \sigma \rangle =0$ and $\langle \mu \rangle
=0$ \cite{km}. According to (\ref{22}), for these kind of systems
(with just one scalar order parameter) we can basically have only
three phases. One with $\langle \sigma \rangle \neq 0$ and $\langle
\mu \rangle =0$, another with $ \langle \sigma \rangle =0$ and
 $\langle \mu \rangle \neq 0$ and finally a phase with both $\langle \sigma \rangle = 0$ and $\langle \mu \rangle =0$.

From the large distance behavior of the quantum topological
correlators, we show that the three phases allowed by the dual
algebra are realized in the disordered system considered below.

\section{The CP$^1$ Formulation of the Short-Range AF Heisenberg Spin-Glass}
\setcounter{equation}{0}
\subsection{The Nonlinear Sigma Model Formulation }

 We review in this and in the next subsection the field-theoretical description of the
 disordered SO(3) quantum Heisenberg-like system, which was studied with great
 detail in \cite{mm,mmartigo}. It is described by the hamiltonian operator,
\begin{equation}
\widehat{\mathcal{H}}=\sum_{\langle ij\rangle} J_{ij}
\mathbf{\widehat{S}}_{i}\cdot\mathbf{\widehat{S}}_{j},\label{Hpuro}
\end{equation}
with nearest-neighbor interactions on a $2D$ square lattice of
spacing $a$, having the random couplings $J_{ij}$ associated with a
Gaussian probability distribution with variance $\Delta J$ and
centered in $\bar{J}>0$, namely,
\begin{equation}
P[J_{ij}]=\frac{1}{\sqrt{2\pi(\Delta
J)^{2}}}\exp\left[-\frac{(J_{ij}-\bar{J})^{2}}{2(\Delta
J)^{2}}\right]\label{distgauss}.
\end{equation}

We consider the case of quenched disorder, which is conveniently
dealt with the replica method \cite{ea,by}. In this case the average
free-energy is given by
\begin{equation}
\overline{F}=-k_{B}T\lim_{n\longrightarrow
0}\frac{1}{n}\left([Z^{n}]_{av}-1\right),\label{fb}
\end{equation}
where $[Z^{n}]_{av}$ is the disorder-averaged replicated partition
function.

With the help of spin coherent states $|\mathbf{\Omega}_{i}^{\alpha}(\tau)\rangle$ such that
\begin{equation}
\langle\mathbf{\Omega}_{i}^{\alpha}(\tau)|\mathbf{\widehat{S}}^\alpha_{i}
|\mathbf{\Omega}_{i}^{\alpha}(\tau)\rangle = S \mathbf{\Omega}_{i}^{\alpha}(\tau),
\end{equation}
where $S$ is the spin quantum number,
it was shown \cite{mm} that in the continuum limit, the average replicated partition function corresponding to
(\ref{Hpuro}) could be written as
\begin{equation}
[Z^{n}]_{av}=\int\mathcal{D}\mathbf{n}\mathcal{D}Q \mathcal{D}\lambda\exp\left\{- \int d\tau  L_{\bar J,\Delta}[\mathbf{n}^{\alpha},Q^{\alpha\beta},\lambda^\alpha]\right\},\label{Z}
\end{equation}
where the corresponding lagrangian density is a generalized relativistic nonlinear sigma model (NLSM)

\begin{eqnarray}
\mathcal{L}_{\bar J,\Delta}&=&\frac{1}{2}|\nabla\mathbf{n}^{\alpha}|^{2}+
\frac{1}{2c^{2}}|\partial_{\tau}\mathbf{n}^{\alpha}|^{2}+i\lambda_{\alpha}(|\mathbf{n}^{\alpha}
|^{2}-\rho_{s}) \nonumber  \\
&&+\frac{D}{2} \int d\tau'
\left[Q_{ab}^{\alpha\beta}(\tau,\tau^{\prime})Q_{ab}^{\alpha\beta}(\tau,\tau^{\prime})\right.\label{acaoparadecompor}  \\
&&-\left.\frac{2}{\rho_{s}}n_{a}^{\alpha}(\tau)Q_{ab}^{\alpha\beta}(\tau,\tau^{\prime})n_{b}^{\beta}(\tau^{\prime})\right].\nonumber
\end{eqnarray}
where $D=S^4(\Delta J)^2/a^2$ ($a$: lattice parameter) and
$\rho_s=S^2 \bar J$. In the above expression, summation on the
replica indices $\alpha, \beta= 1,...,n$ is understood.

 The field $\mathbf{n}^\alpha=(\sigma^\alpha, \vec \pi^\alpha)$ is
the continuum limit of the (staggered) spin $\mathbf{\Omega}^\alpha$ and satisfies the constraint
$\mathbf{n}^\alpha\cdot\mathbf{n}^\alpha= \rho_s $, which is implemented by integration on $\lambda^\alpha$.

Decomposing $Q^{\alpha\beta}$ into replica diagonal and off-diagonal parts,
 \begin{equation}
 Q_{ab}^{\alpha\beta}(\vec r;\tau,\tau')\equiv \delta_{ab}[\delta^{\alpha\beta}\chi(\vec r;\tau,\tau')+
q^{\alpha\beta}(\vec r;\tau,\tau')]
\end{equation}
where $q^{\alpha\beta}=0$ for $\alpha = \beta$, we get

\begin{eqnarray}
\mathcal{L}_{\bar J,\Delta}&=&\frac{1}{2}|\nabla\mathbf{n}^{\alpha}|^{2}+
\frac{1}{2c^{2}}|\partial_{\tau}\mathbf{n}^{\alpha}|^{2}+i\lambda_{\alpha}(|\mathbf{n}^{\alpha}
|^{2}-\rho_{s}) \nonumber  \\
&&+\frac{3D}{2} \int d\tau'
\left[n\chi^{2}(\tau,\tau^{\prime})+ q^{\alpha\beta}(\tau,\tau^{\prime})q^{\alpha\beta}(\tau,\tau^{\prime})\right.\nonumber \\
&&-\left.\frac{D}{\rho_{s}}\mathbf{n}^{\alpha}(\tau)\chi(\tau,\tau^{\prime})
\mathbf{n}^{\alpha}(\tau^{\prime})\right.\nonumber  \\
&&-\left.\frac{D}{\rho_{s}}\mathbf{n}^{\alpha}(\tau)q^{\alpha\beta}(\tau,\tau^{\prime})
\mathbf{n}^{\beta}(\tau^{\prime})\right].\label{acaoparadecompor1}
\end{eqnarray}

This was the starting point for the CP$^1$ formulation, which was
derived in \cite{mmartigo}. In a previous work \cite{mm}, we took a
different path: from (\ref{acaoparadecompor1}), we integrated over
the $\vec \pi$-field and thereby obtained an effective action for
the remaining fields. This allowed the determination of the average
free-energy and, out of this, the phase diagram of the system.
Identical results were found within the CP$^1$ framework
\cite{mmartigo}.

\subsection{The CP$^1$ Formulation}

Introducing the CP$^1$ field as usual, by the relation
\begin{equation}
\mathbf{n}^{\alpha}(\tau) = \frac{1}{\sqrt{\rho_s}}\left[z_i^{*\alpha}(\tau)\mathbf{\sigma}_{ij} z_j^{\alpha}(\tau) \right]\label{nz}
\end{equation}
where the $z_i^\alpha$ field satisfies the constraint
\begin{equation}
|z_1^{\alpha}|^2 + |z_2^{\alpha}|^2 = \rho_s.\label{vinc}
\end{equation}
and using the correspondence
\begin{equation}
\frac{1}{2}|\nabla\mathbf{n}^{\alpha}|^{2}+
\frac{1}{2c^{2}}|\partial_{\tau}\mathbf{n}^{\alpha}|^{2} \Leftrightarrow 2 \sum_{i=1}^2|D_\mu z_i^{\alpha}|^2 \label{01},
\end{equation}
 ($D_\mu = \partial_\mu + i A_\mu$), which comprises a functional
integration over the auxiliary vector field
 $A_\mu$, we obtain \cite{mmartigo}

 %This becomes a dynamic $U(1)$ gauge field when the quantum corrections generated by integration on the $z_i^\alpha$-fields
%are taken into account \cite{witten}.

%Using (\ref{nz}), (\ref{vinc}) and (\ref{01}) in (\ref{Z}), we may express the average replicated partition function in terms of the CP$^1$
%fields as
\begin{equation}
[Z^{n}]_{av}=\int\mathcal{D}z  \mathcal{D}z^* \mathcal{D} A_\mu \mathcal{D}\chi \mathcal{D}q \mathcal{D}\lambda
e^{- S}\label{Z1},
\end{equation}
where $ S\left[z_i^\alpha, z_i^{\alpha *}, A_\mu, \lambda,  \chi(\tau,\tau'), q^{\alpha\beta}(\tau,\tau') \right]$
is the action corresponding to the lagrangian density
$$
\mathcal{L}_{\bar J,\Delta,\mathbf{CP^1}}=2 |D_\mu z_i^{\alpha}|^2 +i\lambda_{\alpha}(|z_i^{\alpha}|^2-\rho_{s})
$$
$$
+\frac{3D}{2} \int d\tau'
\left[n\chi^{2}(\tau,\tau^{\prime})+ q^{\alpha\beta}(\tau,\tau^{\prime})q^{\alpha\beta}(\tau,\tau^{\prime})\right]
$$
$$
+\frac{2D}{\rho^2_{s}} \int d\tau' \left \{[\chi(\tau,\tau^{\prime})][|z_i^{*\alpha}(\tau)|^2 |z_j^{\alpha}(\tau^{\prime})|^2]
\right.
$$
\begin{equation}
-\left.   [z_i^{*\alpha} z_j^{\alpha}(\tau)][\chi(\tau,\tau^{\prime})\delta^{\alpha\beta}+ q^{\alpha\beta}(\tau,\tau^{\prime})]
[z_i^{\beta} z_j^{*\beta}(\tau^{\prime})]
\right \},\label{acaoparadecompor11}
\end{equation}

where summation in $i,j,\alpha,\beta$ is understood.

In \cite{mmartigo}, we have determined the average free-energy, by
expanding the fields around their stationary point in (\ref{Z1}) and
integrating the quadratic quantum fluctuations of the $z_i^\alpha$
fields, namely
$$
S\left[z_i^\alpha, z_i^{\alpha *}, A_\mu, \lambda_\alpha,  \chi,
q^{\alpha\beta} \right] =
 S\left[z_{i,\mathrm{s}}^\alpha, z_{i,\mathrm{s}}^{\alpha *}, A_\mu^\mathrm{s},  m^2,  \chi_{\mathrm{s}}, q_{\mathrm{s}}^{\alpha\beta} \right]
 $$
 \begin{equation}
  + \frac{1}{2}  \int d\tau d\tau' \eta_i^{\alpha *}(\tau) \mathbb{M}^{\alpha\beta}_{ij}(\tau,\tau') \eta_j^{\beta } (\tau') \label{Z2},
\end{equation}
where $\eta_i^{\alpha } = z_i^{\alpha } - z_{i,\mathrm{s}}^{\alpha
}$ and $\mathbb{M}$ is the matrix
\begin{eqnarray}
 \mathbb{M} =
\left(
\begin{array}{c}
\frac{\delta^2 S}{\delta z_i^\alpha(\tau) \delta z_j^{*\beta}(\tau^\prime)} \ \ \ \ \   \frac{\delta^2 S}{\delta z_i^{\alpha}(\tau) \delta z_j^{\beta}(\tau^\prime)} \\
\frac{\delta^2 S}{\delta z_i^{*\alpha}(\tau) \delta
z_j^{*\beta}(\tau^\prime)} \ \ \ \ \   \frac{\delta^2 S}{\delta
z_i^{*\alpha}(\tau) \delta z_j^{\beta}(\tau^\prime)}
\end{array}\right)\;\label{mat},
 \label{m}
 \end{eqnarray}
 with elements taken at the stationary fields.

Inserting (\ref{Z2}) in (\ref{Z1}) we obtain, after integrating over
the $z$-fields,
\begin{equation}
[Z^{n}]_{av}= e^{-
n{S}_{\mathrm{eff}}\left[\sigma_{\mathrm{s}}^\alpha, m^2,
A_\mu^\mathrm{s}, q_{\mathrm{s}}^{\alpha\beta}(\tau-\tau'),
\chi_{\mathrm{s}}(\tau-\tau')\right]},\label{funcaoparticaosigma}
\end{equation}
where
\begin{equation}
 S_{\mathrm{eff}} = S\left[\sigma_{\mathrm{s}}^\alpha, m^2, A_\mu^\mathrm{s}, q_{\mathrm{s}}^{\alpha\beta}, \chi_{\mathrm{s}}\right]
 -\frac{1}{n } \ln \mathop{\mathrm{Det}} \mathbb{M}, \label{seff1}
\end{equation}

 We conclude, because of (\ref{fb}), that
\begin{equation}
\bar F= \frac{1}{\beta}
S_{\mathrm{eff}}\left[\sigma_{\mathrm{s}}^\alpha, m^2,
A_\mu^\mathrm{s}, q_{\mathrm{s}}^{\alpha\beta}(\tau-\tau'),
\chi_{\mathrm{s}}(\tau-\tau')\right].
\end{equation}

In the previous expressions, the subscript $s$ means that the fields
are taken at their stationary values:  $ \lambda_{\mathrm{s}}$ ($m^2
= 2i\lambda_{\mathrm{s}}$ is the spin gap), $A_{\mathrm{s},\mu} =0$,
$\chi_{\mathrm{s}}(\tau-\tau')$ and
$q_{\mathrm{s}}^{\alpha\beta}(\tau-\tau')$.

The staggered magnetization $\sigma_{\mathrm{s}}^\alpha$ is given in
terms of the CP$^1$ fields as
\begin{equation}
\sigma^2_{\mathrm{s}} = \frac{1}{n} \sum_{\alpha=1}^n
\left[|z_{1,\mathrm{s}}^\alpha |^2+|z_{2,\mathrm{s}}^\alpha |^2
\right] \equiv \frac{1}{n} \sum_{\alpha=1}^n \sigma^2_\alpha
\label{magn}.
\end{equation}

From the average free-energy  one can derive the phase diagram of
the system, which is shown in Fig.1 \cite{mm,mmartigo}. This
presents a critical line separating the SG and PM phases, which
starts, at $T=0$, in the quantum critical point
\begin{equation}
\rho_0=\frac{\Lambda}{2\pi}\left[1+\frac{1}{\gamma}\left[1+\frac{1}{2}\ln(1+\gamma)\right]\right],
\label{r0}
\end{equation}
where
\begin{equation}
\gamma=3\pi\left(\frac{ \bar J}{\Delta J}\right)^2 = \frac{3\pi
\rho_s^2 \Lambda^2}{D}
\end{equation}
and $\Lambda=1/a$ is the high-momentum cutoff. For $\rho > \rho_0$,
there is an ordered (AF) N\'eel phase at $T=0$.

The Edwards-Anderson (EA) order parameter, which is used for
detecting the occurrence of a SG phase \cite{ea,by}, is given by
\begin{equation}
 q_{\mathrm{EA}} = T \bar q_0
\end{equation}
where
$$\bar q_0 = \frac{1}{n(n-1)} \sum_{\alpha,\beta}
q^{\alpha\beta} (\omega_n=0).
$$
($\omega_n$ are the Matsubara frequencies).

\begin{figure}[ht]
\centerline {
\includegraphics
[clip,width=0.5\textwidth ,angle=0 ] {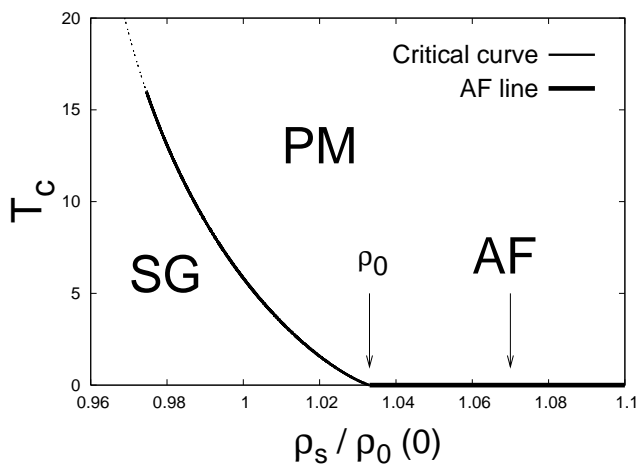} } \caption{Phase
diagram for $\gamma=10^2,\Lambda=10^3$. The critical curve valid
near the QCP $\rho_0$ (solid curve). $\rho_0(0)=\Lambda/2\pi$ is the
QCP of the pure AF system. The value ascribed to $\Lambda$ is a
realistic one in $K$ ($\Lambda \rightarrow \frac{\hbar
v_s}{k_{B}}\Lambda$; $v_s$: spin-wave velocity).} \label{FigRhoxTc}
\end{figure}

A detailed study of the phase diagram of the system, explicitly
showing the $\bar J$ and $\Delta J$ dependence can be found in
\cite{mmartigo}.

We would like to emphasize that in obtaining the above phase
diagram, both in the CP$^1$ and NLSM versions, quantum fluctuations
have been included in the derivation of the average free-energy,
hence it goes beyond the mean-field approximation. The fact that the
ordered AF phase is washed out from any $T \neq 0$ is a clear
evidence for this.

Our extended NLSM depends on the coupling $\bar J$ precisely in the
same way as the usual NLSM. Thus, the analysis of the validity of
the approximations made, as a function of the value of this coupling
follows closely the corresponding analysis well-known in the simple
NLSM \cite{chn}.

\section{ Quantum Skyrmion Correlation Functions}

\setcounter{equation}{0}

\subsection{The General Method}

Given a theory containing an abelian gauge field in two dimensions,
the topological excitations are magnetic vortices
\cite{quantumvort,evora}. Skyrmions, which are topological
excitations of the NLSM, accordingly, appear as magnetic (in the
$A_\mu$-field) vortices in the CP$^1$ formulation. The correlation
functions of the corresponding vortex quantum creation operator
$\mu(\vec x, \tau)$ are obtained by treating this operator as a
disorder variable, dual to the order parameter of the system, as we
saw in section II. Then, a method of quantization has been developed
\cite{quantumvort,evora}, where all correlation functions of the
topological excitation creation operator, can be obtained by
modifying the integrand of the partition function by adding to the
corresponding field intensity tensor $F_{\mu\nu}$, an external
particular field configuration $\tilde B_{\mu\nu}(z;x,y)$. We
implement this method below for the present case.

We will take the CP$^1$ field $z_i$ as the ``order'' field. From
(\ref{magn}), it is clear that whenever $\langle z^\alpha_i\rangle
\neq 0$, the staggered magnetization will not vanish as well. We can
then introduce the dual algebra relating $z_i^\alpha$ with the
topological excitation creation operator, in the form of (\ref{2}).
Using the fact that the symmetry group of the model is U(1), we have
\cite{quantumvort}
\begin{equation}
 \mu(\mathbf{x}, t)  z^\alpha_i (\mathbf{y}, t)  = \exp\left \{i \arg(\mathbf{y} -\mathbf{x})\right \}  z^\alpha_i(\mathbf{y}, t) \mu(\mathbf{x}, t).
\label{4}
\end{equation}
 The vortex creation operator $\mu$ satisfying (\ref{4}) is given by
\cite{quantumvort}
\begin{equation}
 \mu(\mathbf{x}, t)   = \exp\left \{i 2\pi \int d^2\mathbf{r} \arg(\mathbf{r} -\mathbf{x})\sum_{\alpha=1}^n[z^{\alpha*}_i \pi^{\alpha*}_i- z^\alpha_i \pi^{\alpha}_i](\mathbf{r}, t)\right \}  ,
\label{5}
\end{equation}
where $\pi_i^\alpha$ is the momentum canonically conjugate to
$z_i^\alpha$. Whenever the $A_\mu$-field kinematics is described by
a Maxwell term, this can also be written as
$$
 \mu(\mathbf{x}, t)   = \exp\left \{i 2\pi \int d^2\mathbf{r} \arg(\mathbf{r} -\mathbf{x})\partial_i F^{0i}(\mathbf{r}, t)\right \},
$$
by just considering the field equation. Then, using the
Cauchy-Riemann equation for $\arg (\mathbf{r} -\mathbf{x})$ and the
analytical properties of this function, we obtain, equivalently
\begin{equation}
 \mu(\mathbf{x}, t)   = \exp\left \{i 2\pi \int_{(\mathbf{x}, t)}^\infty d\xi_i \epsilon ^{ij} F^{0j}(\mathbf{r}, t)\right \}
\label{6},
\end{equation}
where $F^{\mu\nu}$ is the field intensity tensor corresponding to
$A_\mu$.

This last form of the operator is the most useful for obtaining the
correlation functions. It is not difficult to infer from (\ref{6})
that, for a Maxwell-type action, for which $S_g$ is quadratic in
$F^{\mu\nu}$, the $\mu$-field two-point correlation function will be
given by \cite{quantumvort}
\begin{equation}
\langle \mu(\mathbf{x},\tau) \mu^\dagger(\mathbf{y},\tau') \rangle =
\int \mathcal{D} A_\mu \exp \left \{- S_g \left[ F_{\mu\nu} + \tilde
B_{\mu\nu}(z;x,y)\right ]  \right \} \label{fc1},
\end{equation}
where
\begin{equation}
\tilde B_{\mu\nu}(z;x,y) = 2\pi
\int_{x=(\mathbf{x},\tau)}^{y=(\mathbf{y},\tau')} d\xi_\lambda
\epsilon^{\lambda\mu\nu} \delta (z-\xi) \label{fc2}.
\end{equation}
This turns out to be a general expression for the topological
excitation correlators, namely, it holds for any action $S_g \left[
F_{\mu\nu}\right]$ whatsoever, irrespective of its form
\cite{evora,quantumvort}.

Now, an important adjustment must be made, in order to adapt this
result to the present disordered system. In order to recover the
physical thermodynamics, we must take the limit where the number of
replicas vanishes, $n\rightarrow 0$, hence we must be careful when
defining the physical correlator in this limit. From (\ref{5}), we
see that we actually have
\begin{equation}
\langle \mu \mu^\dagger\rangle  = \prod_{\alpha =1}^n \langle \mu
\mu^\dagger\rangle_\alpha \label{7},
\end{equation}
because fields belonging to different replicas commute. It is
natural, therefore, to define the physical correlation function as
the geometrical average among the $n$ replicas, before taking the
limit $n\rightarrow 0$, namely
\begin{equation}
\langle \mu \mu^\dagger\rangle_{\mathrm{phys}} = \lim_{n\rightarrow
0}\left[\prod_{\alpha =1}^n \langle \mu \mu^\dagger\rangle_\alpha
\right]^{\frac{1}{n}} = \lim_{n\rightarrow 0}\langle \mu
\mu^\dagger\rangle^{\frac{1}{n}} \label{8},
\end{equation}
where the correlator on the r.h.s. is given by (\ref{fc1}).

\subsection{The Effective Gauge Field Theory}

We must now determine the form of the $A_\mu$-field action
$S_g[F_{\mu\nu}]$ in (\ref{fc1}), in order to calculate the
topological quantum correlation function. For this purpose, we start
from the CP$^1$ description of the system, \cite{mmartigo} expand
the action in the $z_i$'s and $A_\mu$ fields up to the second order
around the stationary points ($z_i^\alpha = A_\mu^\mathrm{s}=0$) and
perform the quadratic $z_i$ integrals. We must stress that,
expanding around $z_i^\alpha = 0$, we are only considering the case
when $\langle \sigma \rangle =0$ and therefore this analysis does
not apply to the ordered AF phase. This has been already considered
elsewhere \cite{emqs}.

\begin{equation}
Z_{A_\mu}=\int\mathcal{D}z  \mathcal{D}z^* \mathcal{D}A_\mu e^{-
S\left[z_i^\alpha, z_i^{\alpha *}, A_\mu,  m^2,
\chi_{\mathrm{s}}(\tau-\tau').
q_{\mathrm{s}}^{\alpha\beta}(\tau-\tau') \right]}\label{Z3}.
\end{equation}
We took the fields $\lambda, \chi$ and $q^{\alpha\beta}$ at their
stationary point leaving the integrals in $z_i$ and $A_\mu$. Then,
integrating on the $z_i$-fields and expanding the action in the
$A_\mu$-field up to quadratic fluctuations around the stationary
configuration $A^{\mathrm{s}}_\mu=0$, we get

\begin{equation}
Z_{A_\mu}=\int \mathcal{D}A_\mu e^{- S_g\left[ A_\mu
\right]}\label{Z4},
\end{equation}
where,
\begin{equation}
 S_g[ A_\mu ] =  \frac{n}{2} \int_0^\beta d\tau \int d^2\mathrm{r} A_\mu \left[ \frac{\delta^2  S_{\mathrm{eff}}[A_\mu]}{\delta A_\mu\delta A_\nu}\right ]_{A_\mu=0}
  A_\nu \label{Z5}.
\end{equation}

In Appendix A, we show that the action for the $A_\mu$-field can be
written as

$$
 S_g[ A_\mu ] =  n \int_0^\beta d\tau\int d^2\mathrm{r} \left\{ \frac{\kappa}{4} F_{\mu\nu} F^{\mu\nu}\right.
 $$
\begin{equation}
 \left. + \frac{\alpha}{4} \int_0^\beta d\tau'\int d^2\mathrm{r}'
  F_{\mu\nu} (\mathrm{r},\tau)\tilde\Pi(\mathrm{r},\mathrm{r}';\tau,\tau')  F^{\mu\nu}(\mathrm{r}',\tau')\right \}
 \label{Z6},
\end{equation}
where $\kappa=\frac{m}{4\pi}$, $ \alpha = \bar q_0  \left(
\frac{D}{2\pi \rho_s}\right ) $ and
\begin{equation}
\tilde\Pi(\vec k, \omega_n)= \frac{T}{[|\vec
k|^2+\omega_n^2]}\int_0^1  \frac{dx}{|\vec k|^2 x(1-x) + x\omega_n^2
+m^2} \label{Z7},
\end{equation}

Notice that the $\alpha$-term in the effective action (\ref{Z6}) is
different from zero only in the SG phase, where $\bar q_0 \neq 0$.

\subsection{The Skyrmion Correlation Functions}

We may write (\ref{fc1}) as
$$
\langle \mu(\vec x,\tau) \mu^\dagger(\vec y,\tau') \rangle = \int
\mathcal{D} A_\mu \exp \left \{- \int_0^\beta\int d^2\mathrm{r}
\left[ \frac{n}{2} A_\mu [-\Box \Pi] A_\nu  \right.\right.
$$
\begin{equation}
\left.\left. + \sqrt{n} \partial_\nu \tilde B^{\nu\mu}[\Pi] A_\mu +
\frac{1}{4}\tilde B_{\mu\nu}[\Pi] \tilde B_{\mu\nu}\right] \right
\}, \label{fc3}
\end{equation}
where
\begin{equation}
\Pi = \kappa + \alpha \tilde \Pi \label{fc4}
\end{equation}

Introducing a gauge fixing term, we can integrate (\ref{fc3}) in
$A_\mu$, obtaining
$$
\langle \mu(\vec x,\tau) \mu^\dagger(\vec y,\tau')
\rangle_{\mathrm{phys}} =  \exp \left \{ 2\pi^2 \int_x^y
d\xi_\lambda \int_x^y d\eta_\rho \epsilon^{\lambda \mu \alpha}
\epsilon^{\rho \nu \alpha} \right.
$$
\begin{equation}
\left. \times \partial_\mu \partial'_\nu F(\vec \xi -\vec \eta;
\xi_0-\eta_0) -  \frac{1}{4}\tilde B_{\mu\nu}[\Pi] \tilde B_{\mu\nu}
\right \} \label{fc5},
\end{equation}
where $x=(\mathbf{x},\tau), y=(\mathbf{y},\tau')$,
\begin{equation}
F(\vec x ; \tau) = T \sum_{\omega_n} \int \frac{d^2k}{(2\pi)^2}
F(\vec k,\omega_n) e^{i \vec k\cdot \vec x}
 e^{-i \omega_n \tau},
\label{fc6}
\end{equation}
and
$$
F(\vec k,\omega_n)=\frac{\Pi}{-\Box} = \frac{\kappa}{|\vec
k|^2+\omega_n^2}
$$
\begin{equation}
+\frac{\alpha T}{[|\vec k|^2+\omega_n^2]^2}\int_0^1  \frac{dx}{|\vec
k|^2 x(1-x) + x\omega_n^2 +m^2} \label{fc7}
\end{equation}
%
%\frac{\Pi(\vec k, \omega_n)}{|\vec k|^2+\omega_n^2}=
Using the identity
$$
\epsilon^{\lambda \mu \alpha} \epsilon^{\rho \nu \alpha} =
\delta^{\lambda\rho} \delta^{\mu\nu}- \delta^{\lambda\nu}
\delta^{\mu\rho}
$$
in (\ref{fc5}) we can see that the first term cancels the last one
in (\ref{fc3}) (and in (\ref{fc5})) and the second one gives
\begin{equation}
\langle \mu(\vec x,\tau) \mu^\dagger(\vec y,\tau')
\rangle_{\mathrm{phys}} =  \exp \left \{ 4\pi^2 [F(\vec x -\vec y;
\tau-\tau')- F(\vec \epsilon; 0)] \right \} \label{fc8},
\end{equation}
where $\epsilon$ is a short-distance cutoff. Introducing the
renormalized skyrmion field operator
$$
\mu_R (\vec x,\tau) = \exp \left \{ 2\pi^2 F(\vec \epsilon; 0)\right
\} \mu_ (\vec x,\tau),
$$
we get the renormalized and finite correlation function
\begin{equation}
\langle \mu_R(\vec x,\tau) \mu_R^\dagger(\vec y,\tau')
\rangle_{\mathrm{phys}} =  \exp \left \{ 4\pi^2 F(\vec x -\vec y;
\tau-\tau') \right \} \label{fc88},
\end{equation}

In Appendix B, we calculate the inverse Fourier transforms of the
$\kappa$ and $\alpha$  terms of the function $F$. These allow us to
determine the large distance behavior of the quantum skyrmion
correlation functions (\ref{fc8}).

\section{Large Distance Behavior of Skyrmion Correlators}
\setcounter{equation}{0}

\subsection{N\'eel Phase}

The skyrmion two-point correlation function has been evaluated in
the ordered AF phase, which occurs on the line $T=0 ; \ \rho_s >
\rho_0$ in \cite{emqs}. It presents the following large distance
behavior
\begin{equation}
\langle \mu_R(\vec x, \tau) \mu_R^\dagger(\vec y,
\tau)\rangle_{\mathrm{phys}}^{AF} \stackrel{|\vec x -\vec y|
\rightarrow \infty}\longrightarrow \exp \left \{-   2 \pi \sigma^2
|\vec x - \vec y|\right \}. \label{35}
\end{equation}
where $\sigma$ is the staggered magnetization satisfying $ \sigma^2
= \frac{1}{8} \left[\rho_s-\rho_0\right]$ \cite{mm}. The equation
above implies
\begin{equation}
\langle \mu_R\rangle^{AF} = 0 \label{36},
\end{equation} meaning that
the quantum skyrmion states $|\mu_R\rangle^{AF}$ are orthogonal to
the ground state and are, consequently, nontrivial. The exponential
decay of the skyrmion correlator, conversely, implies the
corresponding quantum excitations (skyrmion) have a gap $E_g =  2
\pi \sigma^2$. Notice that this gap, which may be written as
\begin{equation}
E_g = \frac{\pi}{4} \left[\rho_s-\rho_0\right]
\label{37},
\end{equation}
vanishes as we approach the quantum phase transition to the SG phase
at the quantum critical point $(T=0, \rho_s=\rho_0)$.

\subsection{The Paramagnetic Phase}

In the PM phase, we have both $\sigma = 0$ and $\bar q_0 = 0$ and,
consequently, only the $\kappa$ term of the function $F$ in
(\ref{fc7}) contributes to the skyrmion correlation function
(\ref{fc8}). We calculated this term, for equal times, in Appendix
B. From (\ref{c4}) we  have, at $T=0$,
\begin{equation}
\langle \mu_R(\vec x,\tau) \mu_R^\dagger(\vec y,\tau)
\rangle_{\mathrm{phys}}^{PM} \stackrel{ T \rightarrow 0}
\longrightarrow \exp \left \{ \frac{\pi\kappa}{|\vec x -\vec y|}
\right \} \label{fc10}.
\end{equation}

For an arbitrary temperature, $T$, we obtain the large-distance
behavior (see (\ref{c5}))

$$
\langle \mu_R(\vec x,\tau) \mu_R^\dagger(\vec y,\tau)
\rangle_{\mathrm{phys}}^{PM} \stackrel{|\vec x -\vec y| \rightarrow
\infty}\longrightarrow \exp \left \{ \frac{\pi\kappa}{|\vec x - \vec
y|} \left [ \frac{1}{2} \right.\right.
$$
\begin{equation}
\left.\left.
 + \frac{\pi T |\vec x - \vec y| }{\sqrt{2}} \coth[\sqrt{2}\pi T |\vec x -
\vec y|]\right ]
 \right \} \label{fc9}.
\end{equation}

From the above expression, we may infer that
\begin{equation}
\langle \mu_R\rangle^{PM} = \exp \left \{\frac{\pi^2 \kappa
T}{2\sqrt{2}} \right \} \neq 0 \label{fc10}.
\end{equation}
This nonzero result is the one to be expected in the PM phase where
the topological excitations should not be genuine excitations, due
to the absence of spontaneous symmetry breaking. The above result
just confirms this fact, by stating that the topological excitation
quantum state $|\mu_R\rangle^{PM}$ is not orthogonal to the ground
state, being therefore trivial.

\subsection{The Spin-Glass Phase}

In the SG phase, we have $\sigma = 0$ but now $\bar q_0 \neq 0$.
Then, the $\alpha$ term of the function $F$ in (\ref{fc7}) will
contribute to the skyrmion correlation function  (\ref{fc8}).We have
calculated this term, for large distances and equal times, in
Appendix B. This has a logarithmic behavior and therefore dominates
the large distance behavior of the function $F$. Indeed, according
to (\ref{c7}), we have
\begin{equation}
F(\vec x -\vec y; 0) \stackrel{|\vec x -\vec y| \rightarrow \infty}\longrightarrow
- \frac{\bar q_0 D}{24 \pi^2 \rho_s m^2 } \left[ 1 - \frac{3T^2}{2 m^2}\right]\ln C |\vec x -\vec y|
\label{fc11},
\end{equation}
where $C$ is a constant.

Inserting this result in (\ref{fc88}), we get
\begin{equation}
\langle \mu_R(\vec x,\tau) \mu_R^\dagger(\vec y,\tau)
\rangle_{\mathrm{phys}}^{SG} \stackrel{|\vec x -\vec y| \rightarrow
\infty}\longrightarrow
 \frac{1}{|\vec x -\vec y|^\nu}
\label{fc12},
\end{equation}
where
\begin{equation}
\nu =  \frac{\bar q_0 D}{6 \rho_s m^2} \left[ 1 - \frac{3T^2}{2 m^2}\right]
\label{fc13}.
\end{equation}
In realistic systems, we always have $T^2 \ll m^2$ (typically
$T\simeq 10 K$ and $m\simeq 100 K$, therefore $\nu$ is always
positive in the SG phase. As a consequence of this
\begin{equation}
\langle \mu_R(\vec x,\tau) \mu_R^\dagger(\vec y,\tau)
\rangle_{\mathrm{phys}}^{SG} \stackrel{|\vec x -\vec y| \rightarrow
\infty}\longrightarrow 0 \label{fc14}
\end{equation}
and therefore
\begin{equation}
\langle \mu_R\rangle^{SG} = 0 \label{fc15}.
\end{equation}
This result shows that the quantum skyrmion states are orthogonal to
the ground state in the SG phase, being therefore non-trivial
quantum states. The power-law behavior of their two-point
correlator, however implies that they have a zero excitation gap.
This is a quite interesting result. It reveals the existence of a
new duality relation between the AF and SG phases. The usual
order-disorder duality occurs between the AF and PM phases, namely
the order and disorder parameters, $\langle\sigma\rangle$ and
$\langle\mu\rangle$ are respectively nonzero in each of these two
phases, while the other vanishes.

The duality between the AF and SG phases, for both of which $q_{EA}
\neq 0$, however, is of a different nature. The spin excitations are
gapless whereas the topological ones are gapped in the AF phase and
conversely the spin excitations are gapped whereas the topological
ones are gapless in the SG phase. Our finding of gapless topological
excitations in the SG phase is a key step in establishing this new
duality relation.

\subsection{The Spin-Glass Phase and the BKT Mechanism}

There is also an important fact related to the existence of gapless
topological excitations and their associate power-law correlators in
the SG phase. When using the CP$^1$ language the skyrmions become
vortices. The power-law behavior of their correlation functions is a
clear indication that we have a Berezinskii-Kosterlitz-Thouless
(BKT) two-dimensional system of vortices \cite{bkt}. Indeed, quantum
vortices do have a large-distance power-law decay in the
low-temperature phase, which exists below the critical point in a
BKT system \cite{mond}. This explains the existence of a SG phase at
a finite temperature, as observed in \cite{mm,mmartigo}: it is a BKT
phase supporting gapless vortices. Conversely, the presence of
gapless spin wave excitations in the AF phase, washes this phase
away at any finite temperature through the well-known Mermin-Wagner
mechanism \cite{mw}. This is the explanation for the asymmetry found
between the SG and AF phases of this system, the former persisting
at a finite $T$, whereas the other only remains at $T=0$.

The operation of the BKT mechanism in this system only occurs
because it is possible to use a CP$^1$ description, which presents
gapless vortices in a SG phase. Thereby one can understand why we
can have a stable SG phase in a quantum SO(3) disordered 2D
Heisenberg system, but not in the corresponding Ising system
\cite{ising}. The latter does not allow a CP$^1$ formulation with
the corresponding vortices and therefore cannot display the BKT
mechanism.

\subsection{A New Characterization of the Spin-Glass Phase }

The SG phase is the realization of one of the possible phases
allowed by the dual algebra of spin and topological excitation
operators, namely the one where both $\langle\sigma\rangle=0$ and
$\langle\mu\rangle=0$. This criterion -- both order and disorder
parameters vanishing -- can be used as a more symmetrical
alternative for the characterization of a SG phase then the usual
one where  $\langle\sigma\rangle=0$ and $q_{EA} \neq 0$. We can
state, equivalently that a SG phase is one where the quantum
topological excitations are nontrivial gapless states of the Hilbert
space.

One can speculate whether this is a general property or a
peculiarity of the present model. From the point of view of
energetics, it is clear that the creation of a skyrmion defect out
of an ordered ground state, such as the one we have in an AF phase,
costs a finite amount of energy since a number of spins must be
flipped, in order to create the quantum defect state. Conversely,
creating such a defect on a disordered ground state, as the one we
have in a paramagnetic phase, clearly does not change the state of
the system and therefore costs no energy. The skyrmion operator
actually does not not create a truly new state. We can consider that
the topological defects are condensed in the ground state, thereby
producing the disordered PM state. Skyrmions here are not genuine
excitations.

In a spin glass phase the ground state is also a disordered state.
Therefore, when creating a skyrmion defect on such a state, we may
conclude that, as in the PM phase, there will be no cost in energy,
because of the disordered character of the ground state. However,
differently from the PM phase, the SG ground state is a ``frozen''
disordered state. Consequently, the new state generated by the
inclusion of the skyrmion will be nontrivially different from the
ground state, despite being also disordered and having zero energy
cost for its creation. The skyrmion state, hence, must be orthogonal
to the ground state, implying the correlation function must vanish
at large distances. This way the large distance behavior of the
skyrmion correlation function detects the frozen nature of the
ground state. The result is the occurrence of zero energy nontrivial
topological states orthogonal to the ground state. The fact that
they bear a nonzero topological charge clearly distinguishes them
from the ground state.

These arguments may be generalized for Ising systems, for instance,
by replacing skyrmions by Bloch walls and can be applied whenever
topological defects may be introduced. It seems to allow a broad
characterization of a SG phase, for a vast class of systems, as one
for which both the order and disorder parameters vanish, namely
$\langle \sigma \rangle = 0 $, $\langle \mu \rangle =0 $, where
$\sigma$ is the spin and $\mu$ is the topological excitation
operators.

%%%%%%%%%%%%%%%%%%%%%%%%%%%%%%%%%%%%%%%%%%%%%%%%%%%%%%%%%%%%%%%%%%%%%%%%%%%%%%%%%%%%

\section{Conclusion}

The disordered magnetic system considered in this work presents all
the possible phases allowed by the duality relation, which exists
between the staggered spin operator and the creation operator of
quantum topological excitations, namely quantum skyrmions. There is
an ordered antiferromagnetic phase with $\langle \sigma \rangle \neq
0$, $\langle \mu \rangle = 0$, a paramagnetic phase with $\langle
\sigma \rangle = 0$, $\langle \mu \rangle \neq 0$ and a spin-glass
phase, with $\langle \sigma \rangle = 0$, $\langle \mu \rangle = 0$.

The PM and AF phases are dual to each other, in the sense that the
order and disorder parameters, respectively $\langle \sigma \rangle
$ and $\langle \mu \rangle $ show a complementary behavior, being
zero or not, respectively, in each of the two phases. An interesting
duality relation, however, also exists between the AF and SG phases,
concerning the gap of topological and spin excitations. In the AF
phase, we have gapless spin excitations, namely $m =0$, whereas the
quantum topological excitations have a finite gap proportional to
$\sigma^2$ and given by (\ref{37}). The SG phase, conversely,
presents spin excitations with a non-vanishing gap, namely $m \neq
0$ \cite{mm,mmartigo}, whereas the topological excitations are
gapless, according to (\ref{fc12}). The AF and SG phases, therefore
are dual with respect to the gap of the spin and topological
excitations. The paramagnetic phase presents spin excitations with a
gap $m \neq 0$, but there are no genuine topological excitations in
the Hilbert space, since the creation operator of topological
excitations acting on the ground-state produces basically the same
state.

The fact that in the SG phase the skyrmion state is orthogonal to
the vacuum in spite of having zero energy is an indication that the
ground state is frozen albeit disordered. We, therefore arrive at an
alternative characterization of a spin glass state, as one in which
both the order and disorder parameters vanish and the quantum
topological excitations in a magnetic systems are gapless but
nontrivial.

Our results indicate that the stability of the SG phase, where both
$\langle \sigma \rangle $ and $\langle \mu \rangle $ vanish, at a
finite temperature, is a consequence of the BKT mechanism. This does
not apply to Ising systems, where a vortex picture of topological
excitations does not exist.

The average free-energy of a spin-glass is expected to have a large
amount of local minima. The problem of determining the absolute
minimum is a difficult one. One can find results in the literature,
indicating that renormalization group flows may drive a replica
symmetric solution towards a broken replica symmetry one. One
possibility is that this would happen in a high coupling $(\bar J)$
limit. This would bring the system to a new local minimum where the
replica symmetry would be broken. This is a very interesting subject
for future investigation. It is, however beyond the scope of the
present work.

\begin{acknowledgments}

ECM would like to thank Curt Callan and the Physics Department of
Princeton University, where part of this work was done, for the kind
hospitality. This work was supported in part by CNPq and FAPERJ.
CMSC was supported by FAPERJ. We are grateful to P.R.Wells for the
help with the graphics.

\end{acknowledgments}

\appendix
\renewcommand{\theequation}{\thesection.\arabic{equation}}
\setcounter{equation}{0}

%\section{Appendixes}

\section{Effective Action for $A_\mu$}

Let us derive here the expression for the action of the $A_\mu$-field. We will consider only the case $\sigma =0$, namely, the PM and
SG phases. From (\ref{seff1}) it follows, in this case that

\begin{equation}
\Pi^{\mu\nu} = \left[ \frac{\delta^2  S_{\mathrm{eff}}[A_\mu]}{\delta A_\mu\delta A_\nu}\right ]_{A_\mu =0}=
\frac{1}{n } \frac{\delta^2 }{\delta A_\mu\delta A_\nu} \left [\ln \mathop{\mathrm{Det}} \mathbb{M}[A_\mu]\right ]_{A_\mu =0}.
\label{b1}
\end{equation}

By expanding in $A_\mu$ and taking the derivatives before making the
limit $A_\mu \rightarrow 0$, we find in frequency-momentum space
$k^\mu=(\omega_n, \vec k)$,
\begin{equation}
\Pi^{\mu\nu}(\omega_n, \vec k) = \kappa (k^2 \delta^{\mu\nu} - k^\mu
k^\nu) + A \bar q_0 \Gamma(\omega_n,\vec k) \delta^{\mu\nu},
\label{B2}
\end{equation}
where
$$
\Gamma(\omega_n, \vec k) = T \int\frac{d^{2}q}{(2\pi)^{2}}
\frac{1}{[|\vec q|^2 + m^2][|\vec k -\vec q|^2 + \omega^2_n + m^2]}
$$
\begin{equation}
= \frac{T}{4\pi}\int_0^1  \frac{dx}{|\vec k|^2 x(1-x) + x\omega_n^2
+m^2} \label{b3}
\end{equation}

Multiplying and dividing the last term in (\ref{B2}) by $k^2 =
(|\vec k |^2 + \omega^2_n)$ and  adding a pure gauge term, the
result (\ref{Z6}) follows, with
$$
\tilde \Pi =\frac{\Gamma}{k^2}.
 $$

\section{Finite Temperature Inverse Fourier Transforms}

\setcounter{equation}{0}

\subsection{The Inverse Transform of $\frac{1}{k^2+\omega_n^2}$}

We have $\mathcal{F}^{-1}\left [ \frac{1}{k^2+\omega_n^2} \right ]$ given by
\begin{equation}
\mathcal{F}_1(\vec x ; \tau) =T \sum_{\omega_n}\int \frac{d^{2}k}{(2\pi)^{2}}
\frac{e^{i (\vec k \cdot \vec x- i \omega_n \tau )}  }{k^2+\omega_n^2}
 \label{c1}
\end{equation}
The $\vec k$-integral may be easily done \cite{gr}, yielding
$$
\mathcal{F}_1(\vec x ; \tau) = \frac{T}{2\pi}\left \{ -
\lim_{\epsilon \rightarrow 0} \ln \frac{\epsilon}{2}|\vec x|\right.
$$
\begin{equation}
\left. + 2 \sum_{n=1}^\infty K_0(|\omega_n| |\vec x|) e^{- i
\omega_n \tau }  \right \},
 \label{c2}
\end{equation}
 where $K_0$ is a modified Bessel function.
For $\tau =0$ we can perform the sum \cite{gr}. The logarithmic term
is canceled and we get, up to a constant (which will not contribute
to the skyrmion correlation function, as we can infer from
(\ref{fc8}))
$$
\mathcal{F}_1(\vec x ; 0)  = \frac{1}{4\pi |\vec x - \vec y|} \left \{ 1 \right.
$$
\begin{equation}
\left.
+ T  \sum_{n=1}^{\infty} \left[
\frac{1}{\sqrt{ T^2  +  \frac{n^2}{|\vec x - \vec y|^2}}}-\frac{1}{\sqrt{\frac{n^2}{|\vec x - \vec y|^2}}} \right ]\right \}.
\label{c3}
\end{equation}

We immediately see that, for zero temperature we get
\begin{equation}
\mathcal{F}_1(\vec x ; 0)   \stackrel{T \rightarrow
0}\longrightarrow \frac{1}{4\pi |\vec x - \vec y|} . \label{c4}
\end{equation}

For finite temperatures, we obtain from (\ref{c3}), at large
distances
$$
 \mathcal{F}_1(\vec x ; 0)
\stackrel{|\vec x - \vec y| \rightarrow \infty}\longrightarrow
\frac{1}{4\pi |\vec x - \vec y|} \left \{ \frac{1}{2} \right.
$$
\begin{equation}
\left.
 + \frac{\pi T |\vec x - \vec y| }{\sqrt{2}} \coth[\sqrt{2}\pi T |\vec x -
\vec y|]\right \} . \label{c5}
\end{equation}

\subsection{The Inverse Transform of $\frac{\tilde \Pi(\vec k,\omega_n)}{[k^2+\omega_n^2]^2}$}

Now we have $\mathcal{F}^{-1}\left [ \frac{\tilde \Pi(\vec k,\omega_n)}{[k^2+\omega_n^2]^2} \right ]$ given by
\begin{equation}
\mathcal{F}_2(\vec x ; \tau) =T \sum_{\omega_n}\int \frac{d^{2}k}{(2\pi)^{2}}
\frac{\tilde\Pi(|\vec k|,\omega_n)}{[k^2+\omega_n^2]^2} e^{i (\vec k \cdot \vec x- i \omega_n \tau )}
 \label{c4}
\end{equation}
Integrating on the angular $\vec k$-variable, we get
$$
\mathcal{F}_2(\vec x ; \tau) =\frac{T}{4\pi}
\sum_{\omega_n}\int_0^\infty dk \left[
\frac{-1}{2}\frac{\partial}{\partial k} \left (
\frac{1}{k^2+\omega_n^2}\right ) \right ] \tilde\Pi(|\vec
k|,\omega_n)
$$
\begin{equation}
\times J_0(k|\vec x|) e^{- i \omega_n \tau }
 \label{c5}
\end{equation}
 We are actually interested in the large distance behavior of
$\mathcal{F}_2(\vec x ; 0)$. In this regime, only the $k \rightarrow
0$ will contribute to the $k$-integral. Since $\tilde\Pi(|\vec
k|,\omega_n)$ is always regular for $k \rightarrow 0$, we can
replace it by $\tilde\Pi(0,\omega_n)$. Then the $\omega_n$ sum may
be performed, for $\tau =0$, yielding to leading order in $k$:
\begin{equation}
\mathcal{F}_2(\vec x ; 0)  \stackrel{|\vec x | \rightarrow \infty}\longrightarrow
\frac{1}{12\pi m^2 }\left[1 - \frac{3 T^2}{2 m^2} \right ] \lim_{\epsilon \rightarrow 0}\int_0^\infty dk k \frac{J_0(k|\vec x|)}{k^2+\epsilon^2}.
 \label{c6}
\end{equation}
The last integral gives \cite{gr} $K_0(\epsilon |\vec x |) \stackrel{\epsilon \rightarrow 0}\longrightarrow - \ln C |\vec x |$, hence
\begin{equation}
\mathcal{F}_2(\vec x ; 0)   \stackrel{|\vec x | \rightarrow \infty}\longrightarrow   - \frac{1}{12\pi m^2 } \left[1 - \frac{3 T^2}{2 m^2} \right ]
 \ln C |\vec x |
 \label{c7}
\end{equation}
This immediately leads to (\ref{fc11}) and to the power law behavior of the skyrmion correlation function in the SG phase, (\ref{fc12}) and (\ref{fc13}). The constant $C$ will not contribute to the skyrmion correlation function, as we can infer from (\ref{fc8}) and therefore is
not important in this framework.

%%%%%%%%%%%%%%%%%%%%%%%%%%%%%%%%%%%%%%%%%%%%%%%%%%%%%%%%%%%%%%%%%%%%%%%%%%%%%%%%%%%%%%%%%%%%%%%%%%%%

\bibliography{apssamp}% Produces the bibliography via BibTeX.

\end{document}